\documentclass[journal = ancac3,manuscript=article]{achemso}
\setkeys{acs}{articletitle = true}
\usepackage[utf8]{inputenc}
\usepackage{gensymb}
\usepackage{chemformula}
\usepackage{ulem}
\usepackage{graphicx}
\graphicspath{ {./figures/} }

\author{F\'elix Thouin}
\affiliation{Department of Engineering Physics, \'Ecole Polytechnique de Montr\'eal, Montr\'eal, H3T 1J4, QC, Canada}
\altaffiliation{Current address: Department of Physics and Institut Courtois, Universit\'e de Montr\'eal, Montr\'eal, H2V 0B3, QC, Canada}
\author{David M. Myers$^*$}
\affiliation{Department of Engineering Physics, \'Ecole Polytechnique de Montr\'eal, Montr\'eal, H3T 1J4, QC, Canada}
\author{Ashutosh Patri$^*$}
\affiliation{Department of Electrical Engineering , \'Ecole Polytechnique de Montr\'eal, Montr\'eal, H3T 1J4, QC, Canada}
\author{Bill Baloukas}
\affiliation{Department of Engineering Physics, \'Ecole Polytechnique de Montr\'eal, Montr\'eal, H3T 1J4, QC, Canada}
\author{Ludvik Martinu}
\affiliation{Department of Engineering Physics, \'Ecole Polytechnique de Montr\'eal, Montr\'eal, H3T 1J4, QC, Canada}
\author{Antonio I. Fern\'andez-Dom\'inguez}
\affiliation{Departamento de Física Teórica de la Materia Condensada and Condensed Matter Physics Center (IFIMAC),
Universidad Autónoma de Madrid, E-28049 Madrid, Spain}
\author{St\'ephane K\'ena-Cohen}
\email{s.kena-cohen@polymtl.ca}
\affiliation{Department of Engineering Physics, \'Ecole Polytechnique de Montr\'eal, Montr\'eal, H3T 1J4, QC, Canada}

\title{Field-enhancement and nonlocal effects in epsilon-near-zero photonic gap antennas}
\date{December 2021}
\SectionNumbersOff
\abbreviations{ENZ,PGA,ITO,THG,DFSS}
\begin{document}
\maketitle
\begin{abstract}
In recent years, the large electric field enhancement and tight spatial confinement supported by the so-called epsilon near-zero (ENZ) mode has attracted significant attention for the realization of efficient nonlinear optical devices. Here, we experimentally demonstrate ENZ photonic gap antennas (PGAs), which consist of a dielectric pillar within which a thin slab of indium tin oxide (ITO) material is embedded. In ENZ PGAs, hybrid dielectric-ENZ modes emerge from strong coupling between the dielectric antenna modes and the ENZ bulk plasmon resonance. These hybrid modes efficiently couple to free space and allow for large enhancements of the incident electric field over nearly an octave bandwidth, without the stringent lateral nanofabrication requirements of conventional plasmonic or dielectric nanoantennas. To understand the modal features, we probe the linear response of single ENZ PGAs with dark field scattering and interpret the results in terms of a simple coupled oscillator framework. Third harmonic generation (THG) is used to probe the ITO local fields and large enhancements are observed in the THG efficiency over a broad spectral range. Surprisingly, sharp peaks emerge on top of the nonlinear response, which were not predicted by full wave calculations. These peaks are attributed to the ENZ material's nonlocal response, which once included using a hydrodynamic model for the ITO permittivity improves the agreement of our calculations for both the linear and nonlinear response. This proof of concept demonstrates the potential of ENZ PGAs, which we have previously shown can support electric field enhancements of up to 100--200X, and the importance of including nonlocal effects when describing the response of thin ENZ layers. Importantly, inclusion of the ITO nonlocality leads to increases in the predicted field enhancement, as compared to the local calculation.
\end{abstract}

\section{Introduction}

Advances in optical signal processing, optical neural networks and photonic quantum computing are urgently driving the need for nonlinear optical components with ever smaller footprints and lower energy consumption. In this context, epsilon-near-zero (ENZ) materials have attracted significant attention. In an ENZ material, the real part of the permittivity crosses zero due to the presence of a natural (e.g. plasma, phonon, exciton) or artificial resonance (e.g. metamaterial).  This results in exotic optical phenomena such as infinite phase velocity\cite{edwards_experimental_2008} and vanishing photonic density of states\cite{lobet_fundamental_2020}.  Notably, thin films of ENZ materials have been reported to have exceptionally high optical nonlinearities\cite{passler_second_2019,reshef_nonlinear_2019,wen_doubly_2018}. Doped semiconductors are particularly interesting, given that their plasma frequency can be readily controlled via the dopant concentration and tuned in-situ through electrostatic gating. To a large extent, however, the nonlinearities reported in ENZ materials are a consequence of the field enhancement intrinsically provided by ENZ thin films \cite{vassant_berreman_2012,campione_theory_2015}. Importantly, these support so-called Berreman modes, which can be directly excited within the light cone using TM-polarized light. These lead to modest enhancements of the incident electric field for high angles of incidence that can nevertheless strongly enhance high-order nonlinear processes \cite{alam_large_2018}. For a thin indium tin oxide (ITO) layer serving as the ENZ medium, for example, a typical enhancement$E_{\rm ITO}/E_0 \sim 2-3$ can be anticipated\cite{campione_epsilon-near-zero_2015,alam_large_2018}, where $E_0$ and $E_{\rm ITO}$ are the magnitudes of the electric field in air and ITO, respectively. In addition, ultrathin films support a so-called ENZ mode, which is nearly dispersionless and allows for deep sub-wavelength confinement\cite{campione_theory_2015}. This mode can show much larger field enhancements (typ. $E_{\rm ITO}/E_0 \sim 10-20$), but it exists beyond the light cone and can only be excited using e.g. a Kretschmann configuration (ideally at the critical coupling angle) or some form of mode matching with free space.\cite{luk_enhanced_2015}. In both cases, enhancements typically occur over a 100--200 nm bandwidth around the ENZ frequency. Many experiments have explored such effects by placing either metal \cite{alam_large_2018,hendrickson_coupling_2018,kim_role_2016,schulz_optical_2016,campione_near-infrared_2016,jun_epsilon-near-zero_2013,Deng2020,Dass2020,Bruno2020} or dielectric \cite{wang_large_2022} optical antennas on ENZ thin films to increase coupling with free space, provide further lateral confinement or engineer the spectral response.

We have previously reported that all-dielectric photonic gap antennas (PGAs) consisting of a thin low-index medium, sandwiched in a high index dielectric pillar, can provide extremely large Purcell factors, strong directionality and large field enhancements over a broad spectral range\cite{patri_photonic_2021}. These rely principally on the electric field enhancement ensured by the continuity of the displacement field across the interface between high ($\epsilon_{\rm high}$) and low permittivity ($\epsilon_{\rm low}$) media $\epsilon_{\rm low}E_{\rm low}=\epsilon_{\rm high}E_{\rm high}$, where $E_{\rm  low/high}$ are the normal components of the electric field near the interface. Photonic gap antennas also have the significant advantage of not requiring any deep sub-wavelength lithography as is typically required for most optical nanoantennas. They instead simply rely on the fabrication of a very thin low-index layer. In the extreme case, where the thin layer is an ENZ material $\epsilon_{\rm low}\sim 0$, we have shown that the bulk plasmon mode hybridizes with the bare modes of the all-dielectric antenna, leading to hybrid modes with extremely large field enhancements.\cite{patri_hybrid_2022} Full wave electromagnetic calculations show that such structures can provide field enhancements of $\sim 100$ using 2 nm-thick films of ITO and $\sim 150$ in the mid-infrared using 20 nm-thick doped GaAs, as the ENZ media.

In this letter, we report on the design, fabrication and characterisation of dielectric PGAs hosting a 10 nm-thick layer of ITO as the ENZ material. Using dark field scattering spectroscopy, we observe several resonances in the infrared. Frequency domain finite-element method simulations of the far-field scattering allow us to assign these to the different hybrid polaritonic modes supported by the PGA. However, small differences in scattering intensity and resonance wavelength are observed in the calculations, as compared to the experiment. To probe the electric field enhancement inside the ENZ layer, we measure third harmonic generation (THG). The THG shows a strong response spanning $>500$~nm, including strong resonances, which were not anticipated from full-wave calculations. We find that including nonlocal effects in the ITO permittivity, within the framework of the hydrodynamic model, improves the agreement of our scattering calculations and predicts the emergence of the experimentally observed sharp THG resonances. Importantly, the inclusion of nonlocal effects predicts stronger field enhancement than that using a purely local dielectric response, due to a smearing of the induced surface charges at the ITO boundaries~\cite{Wiener2012}. Peak instantaneous THG efficiencies of about 100 $\mu$m$^4$/MW$^2$ are measured at the THG resonances. Despite absorption losses in the antenna's dielectric layers, these efficiencies exceed those recently reported of for mutliresonnant Ge nanodisk antennas\cite{grinblat_degenerate_2017} and could be significantly improved through the use of thinner ITO layers.\cite{Datta2020}

\section{Results and discussion}

\begin{figure}
    \centering
    \includegraphics{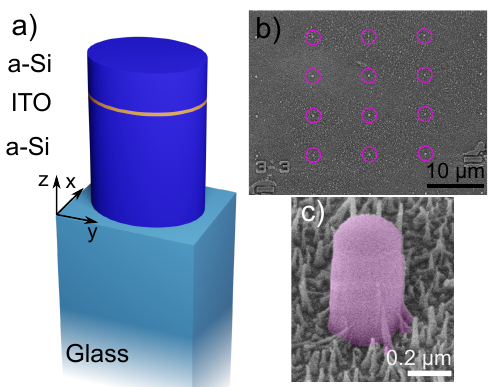}
    \caption{Design and dimensions of the photonic gap antennas. a) A render of the PGA design. It consists of amorphous silicon pillars of elliptical cross section in which a thin ITO layer is embedded. The antennas stand on a thick glass substrate. The thickness of the glass/a-Si/ITO/a-Si layers are 2\;mm, 373\;nm, 10\;nm and 128\;nm respectively. b) SEM image of an antenna array on the sample. The antennas are encircled. c) A close-up on a single antenna, highlighted for clarity. The fabricated antennas are faithful to the design despite thin grass-like structures around them and slightly bent sidewalls.}
    \label{fig:sem}
\end{figure}

A schematic of the fabricated PGA is depicted in Figure \ref{fig:sem} a).
It consists of a high refractive index pillar of elliptical horizontal cross section in which a thin layer of ENZ material is embedded.
The design stems from considerations discussed in a previous publication \cite{patri_photonic_2021}.
The ENZ layer is placed at three quarters of the pillar's height to improve the directivity of the antenna through modal interference.
The pillar's elliptical cross section is chosen to improve the antenna's field enhancement by concentrating the light intensity midway through the pillar's major radius. For operation in the near infrared, amorphous silicon (a-Si) and ITO are chosen as the high refractive index and ENZ materials respectively. Indium tin oxide, which is a conductive oxide commonly used as a transparent conductor, has an ENZ frequency in this range. The exact crossing point of the real part of the permittivity, which is a consequence of the plasma frequency, and thus the carrier concentration, can be strongly tuned through the oxygen vacancy density\cite{cleary_optical_2018}.
In our highly conductive samples, the ITO ENZ wavelength was measured to be 1188\;nm using variable angle spectroscopic ellipsometry (see Supporting Figure S1) performed on a 19 nm-thick film grown under the same conditions as those used in our PGA (see Supporting Information).

The fabricated elliptical a-Si/ITO/a-Si PGAs have layer thicknesses of 373\;nm, 10\;nm and 128\;nm, respectively. While the predicted field enhancement could be further increased for even thinner ITO films, 10 nm was chosen as it allows for relatively uniform films to be achieved directly using reactive sputtering under a wide range of conditions (see Supporting Information Figure S3). Achieving high-quality, thinner (down to monolayer) films is possible, but requires substantial optimization of the deposition conditions or specialized printing techniques.\cite{Li2019,Datta2020} Antennas were patterned laterally using electron beam lithography (see Methods). Each fabricated array holds nominally identical pillars spaced 10\;$\mu$m apart to avoid coupling between them.
To investigate the impact of the antenna dimensions on the response, three arrays (I, II and III) hosting antennas with different cross-sectional dimensions were investigated.
Scanning electron microscope (SEM) micrographs of one of these arrays as well as a close up on a single antenna are shown in Figures \ref{fig:sem} b) and c) respectively.
The cross-sectional dimensions of the antennas in these arrays are summarized in Table\;\ref{tab:pillardimensions}.
These were measured using top-down SEM micrographs shown in Figure S4 of the Supporting Information.
These micrographs confirm that the fabricated structures are faithful to the design, but with some minor imperfections such as a slightly narrower top and the presence of micrograss between the antennas.
The latter occurs due to the random formation of silicon oxyfluoride micromasks during the etching of the bottom a-Si layer.\cite{jansen_black_1995}
However, as we will demonstrate with further optical characterisation, these have no measurable impact on the antenna response.

\begin{table}
    \centering
    \begin{tabular}{c|c|c}
    Array index & Minor diameter (nm) & Major diameter (nm)  \\
    \hline
    I&$230\pm10$&$290\pm10$\\
II&$265\pm5$&$320\pm7$\\
III&$290\pm10$&$365\pm5$\\
    \end{tabular}
    \caption{Cross-sectional dimensions of the investigated antennas}
    \label{tab:pillardimensions}
\end{table}

\begin{figure}
    \centering
    \includegraphics{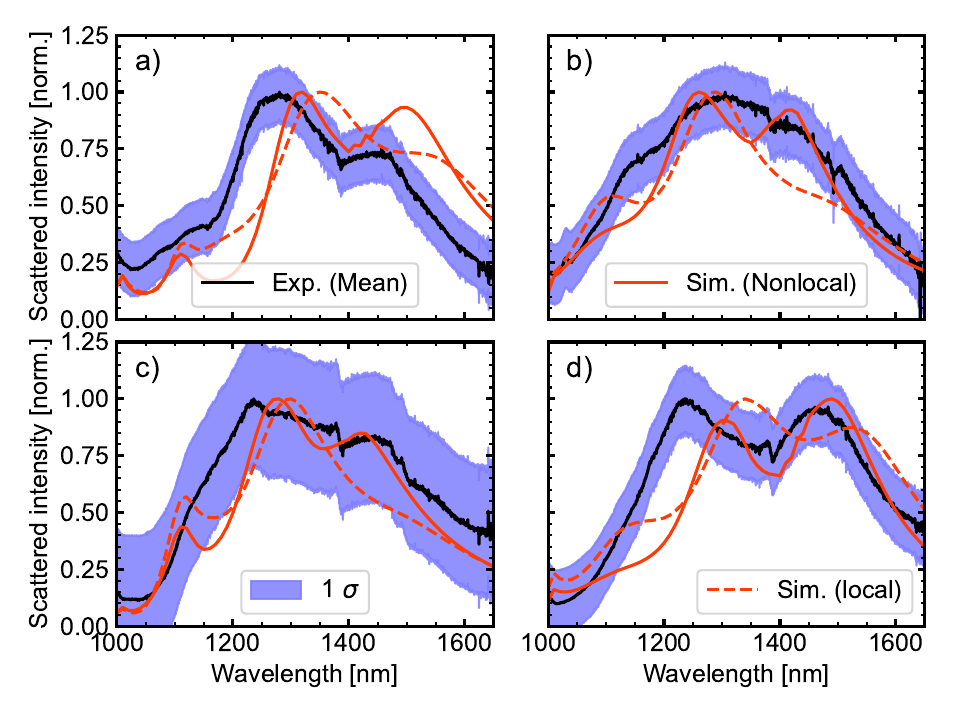}
    \caption{Experimental and simulated DFS spectra of antennas in array II when exciting with transverse electric (a, c) or magnetic (b,d) incident polarisation. The probe light was incident along the minor (a, b) or major (c,d) axes of the antennas' cross-section. The experimental spectra are averaged over those of many individual antennas (black line, blue areas correspond to a standard deviation) and compared to local (dashed red line) and nonlocal (full red line) simulations.}
    \label{fig:dfss}
\end{figure}

To probe their linear optical response, we first measure the far-field scattering spectrum of individual antennas in array II using polarised single particle dark field scattering (DFS) spectroscopy.
The average DFS spectra are plotted in Figure \ref{fig:dfss} for a transverse electric (a, c) or magnetic (b, d) polarized probe incident along the minor (a, b) or major (c, d) axes of the antenna cross sections.
The DFS spectra of other antennas within the array are qualitatively similar with slight variations in peak intensities and position due to structural inhomogeneities.
The impact of these variations on the scattering cross section are shown in Figure \ref{fig:dfss} as the purple areas corresponding to a standard deviation around the average (black lines).
Despite these inhomogeneities, overlapping resonances in the scattering cross-section are observed for all probe configurations.
The average unpolarized DFS spectra of arrays I, II and III are shown in Figure S5 of the Supporting Information as well as those of individual antennas in Figure S6.
Their broadband DFS response covers nearly an octave and redshifts as the lateral dimensions of the antennas are increased from array I to III.
The micrograss surrounding the antennas do not lead to any observable dark field scattering in the spectral region probed by our experiment (see Figure S7).\

In Figure \ref{fig:dfss}, we compare these data with the simulated scattering cross-section integrated over the collection objective's numerical aperture.
The dimensions reported in Table \ref{tab:pillardimensions} were used for these calculations.
Assuming a local dielectric response as extracted from spectroscopic ellipsometry, the finite element modelling of an antenna's scattering cross-section is plotted as red dashed lines in Figure \ref{fig:dfss} for different incident polarisation and directions.
The simulation capture the main features of the experimentally measured cross sections, such as the presence of resonances as well as their spectral position and width. As described in Ref.~\citenum{patri_hybrid_2022}, the broadband response arises due to the extra resonance arising from mode hybridization and dissipation in the ITO, which leads to spectrally overlapping resonances.
To the best of our abilities, accounting for the impact of surface roughness, variations in the ITO layer's refractive index and uncertainties on the measured antenna dimensions did not improve the agreement between simulations and experiment.

The use of a local dielectric function implicitly neglects the role of electron-electron interactions in the permittivity of the ENZ layer, which arise mainly from Pauli repulsion occurring on a scale given by the Thomas-Fermi length~\cite{ciraci_probing_2012}. This is of the order of 3 nm in ITO~\cite{Scalora2020,Farid2021}, comparable to the thickness of the ENZ layer in our PGA samples. The fermionic and repulsive character of conduction electrons in ITO will give rise to nonlocal effects in their infrared response, such as the emergence of longitudinal plasmon resonances as recently reported in the ENZ response of doped CdO\cite{de_ceglia_viscoelastic_2018}. To include the contribution of nonlocal effects within the framework of the hydrodynamic model for the ITO permittivity, we use the approach of Luo et al.~\cite{Luo2013}, with a Fermi velocity for ITO~\cite{Farid2021} corresponding to $\beta=1.7\cdot10^6$~m/s. We ignore any viscoelastic contribution to $\beta$.~\cite{de_ceglia_viscoelastic_2018} The nonlocal calculations are shown as red solid lines in Figure \ref{fig:dfss}. We find that most resonance positions are blue-shifted, compared to our local calculations, bringing them in closer agreement with our experimentally measured spectra. In addition, including nonlocal effects greatly improves the experimental agreement of the relative peak intensities for the cases where the electric field is oriented along the minor axis of the PGA (panels b and c).

\begin{figure}
    \centering
    \includegraphics{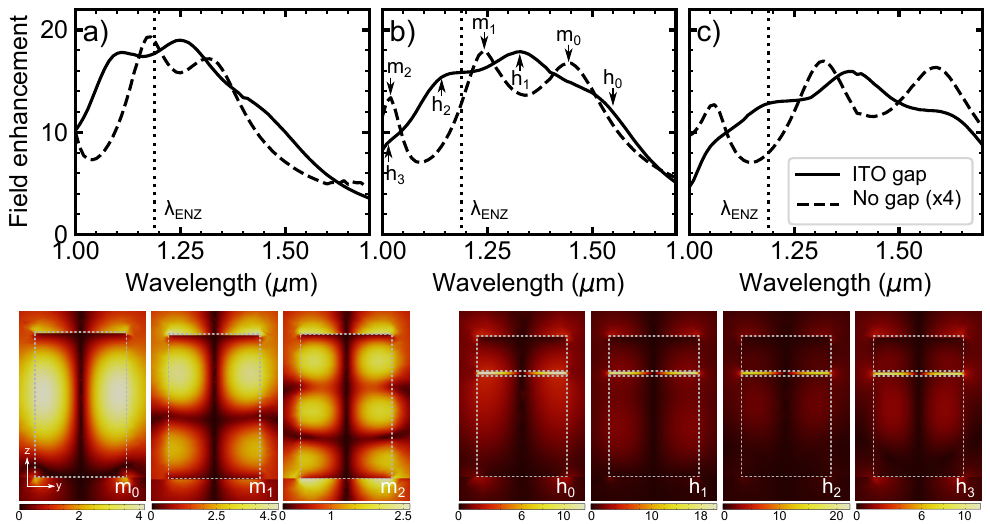}
    \caption{Local electromagnetic simulations of the field enhancement for the antennas in arrays I, II and III, in a), b) and c) respectively, with (full lines) or without (dashed lines) an ITO gap. The dotted line indicates the ENZ wavelength. The lower panels show the magnitude of the z-component of the electric field, $|E_z|/|E_0|$,  under normal excitation for the modes labelled in b). These cuts are taken in the major axis plane of antenna II. The dotted lines outline the boundaries of the antenna.
    }\label{fig:fieldenh}
\end{figure}

The agreement between the experimental DFS spectra and the simulated ones allows us to interpret the behavior of the PGAs by means of the quasinormal modes they support.\cite{patri_hybrid_2022} For this, Figure \ref{fig:fieldenh}a-c) shows the calculated field enhancement spectra of the PGAs, which are more representative of the near-field properties than DFS. We define the field enhancement as the maximum ratio between magnitude of the electric field within the PGA to the magnitude of the incident electric field. The figure also shows the case of ungapped antennas to highlight the effect of the interaction between the ITO slab and the bare dielectric antenna modes.
The PGAs show very distinct resonances within the spectral region shown in the Figure. Like the DFS spectra, when the cross-sectional dimensions of the antennas are increased from Figure \ref{fig:fieldenh} a) to c), these resonances redshift and spread apart.
Compared to the purely dielectric antennas, the ENZ PGA field enhancement spectra exhibit broader, shifted features, an additional resonance, and a fourfold increase in field enhancement.
This behavior is a sign of the strong coupling between the bare dielectric antenna modes and the ITO plasmonic resonance at the ENZ wavelength~\cite{patri_hybrid_2022}.

In the lower panels of Figure~\ref{fig:fieldenh}, the magnitude of the electric field for the resonances of antenna II are shown.
Dielectric modes lead to resonances in the ungapped antenna field enhancement spectra such as those labelled as m$_0$, m$_1$ and m$_2$ in Figure \ref{fig:fieldenh} b).
The corresponding field patterns reveal an increasing number of nodes along the vertical axis for larger mode indices. Meanwhile, the phase of Ez (not shown) exhibits odd parity along the y axis. The strength of the interaction between the ENZ plasmon resonance and a given dielectric mode is dictated by the spatial distribution of its electric field.
By placing the ITO layer at a quarter of the PGA height from the top, the ENZ mode couples most strongly to mode m$_1$, then m$_0$ and least to m$_2$.
This coupling creates new modes labelled h$_0$, h$_1$, h$_2$ and h$_3$ in Figure \ref{fig:fieldenh} b).
These are shifted from the bare antenna resonances and broadened.
These hybrid light-matter modes inherit the efficient near-to-far-field coupling of the bare pillar modes and the strongly localized character of the ITO plasmonic resonance.
These characteristics are apparent in the mode-labelled electric field profiles in the lower panels of Figure \ref{fig:fieldenh}.
All profiles show both a strong localization of the electric field within the ITO layer and a radiative component around the antenna.
This ultimately manifests itself in the broadband field enhancements in Figure \ref{fig:fieldenh}.
It must be noted, however, that some modes do not couple strongly to the ENZ layer.
This is the case for mode h$_3$ of antenna II, in which the ITO layer is located close to a vertical node of m$_2$, leading to a weak coupling with the ENZ resonance.

\begin{figure}
    \centering
    \includegraphics{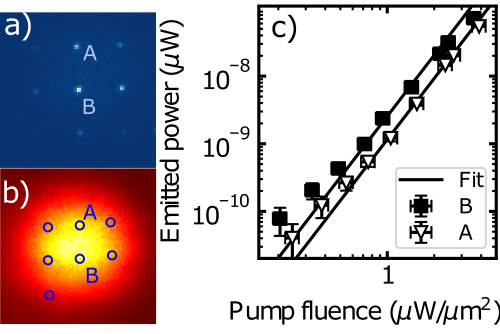}
    \caption{a) Third harmonic image of PGA array II illuminated by a pulsed infrared pump at 1.3\;$\mu$m.
    Antennas are identified through their bright emission.
    Two of them are labelled A and B.
    b) Pump profile at the sample plane used to measure the pump fluence at the location of each antenna.
    Circles are positioned at the corresponding PGA locations extracted from the third harmonic image in a).
    The location of antennas A and B is labelled accordingly.
    c) Third harmonic power emitted from A (triangles) and B (squares) for different pump fluences and their fits to a cubic dependence on pump fluence (solid lines).
    Vertical error bars represent the contribution of shot noise and background noise while horizontal error bars represent uncertainties on the pump power during the measurement.}
    \label{fig:thgpower}
\end{figure}

To experimentally investigate the field enhancement occurring within the fabricated devices, we use THG.
As any non-linear optical phenomenon, THG is a highly sensitive probe of the local field intensity created by a pump beam. As such, it has been widely used to probe field enhancements in metallic\cite{hanke_efficient_2009} and dielectric\cite{xu_boosting_2018} optical nanoantennas, thin films of ENZ materials\cite{capretti_enhanced_2015,luk_enhanced_2015} and metasurfaces\cite{tong_enhanced_2016}.
Our experimental setup allows us to image emission from the sample at the pump's third harmonic and resolve the contribution of individual PGAs (see Methods section for more details).
An image of antenna array II illuminated by a 1.3\;$\mu$m pump, taken at the THG frequency, is shown in Figure\;\ref{fig:thgpower} a).
Intense emission is clearly visible at the antenna positions.
No signal above the background is observed between the antennas, confirming the large field enhancement and the negligible role played by the grass-like background.
To confirm that this emission stems indeed from THG, we check that it follows a cubic dependence on pump fluence.
We compensate for variations in the pump profile over an array by acquiring an image of the pump spot at the sample such as the one shown in Figure\;\ref{fig:thgpower} b) (see Methods for details).
The power emitted from two antennas (A and B in Figure\;\ref{fig:thgpower}) are plotted against pump fluence in Figure\;\ref{fig:thgpower} c).
Both datasets are well fitted over three ordes of magnitude by a cubic dependence on pump fluence, confirming that the observed emission stems from THG inside the antenna.
We cannot discard, however, the contribution of cascaded sum frequency processes to this emission\cite{celebrano_evidence_2019}.
Nevertheless, cascaded or not, THG remains a sensitive probe of field enhancement inside the antennas.

Despite being nominally identical, antenna B is about twice as efficient as antenna A. This reflects the presence of important structural differences between the PGAs to which neither SEM nor DFS spectroscopy are sensitive to. Structural differences between PGAs are unlikely to arise from the bulk of the silicon layers due to their amorphous structure, or from the antenna surfaces given their smooth profiles (see Figure S2b of Supporting Information).
On the other hand, the polycrystalline ITO layer is noticeably rough and features clear crystalline domains of tens of nanometers in size (see Figure S2a of Supporting Information).
When all but the PGA cross sections are etched away from the a-Si/ITO/a-Si multilayer, each antenna samples a different region of the inhomogeneous ITO layer.
Since the PGA radiative properties depend on the overall mode profile, the DFS spectra are only weakly affected by variations in the microstructure.
However, THG is highly sensitive to the permittivity at the electric field hotspots, which are strongly confined to the ITO, as shown in the panels of Figure\;\ref{fig:fieldenh}.

\begin{figure}
    \centering
    \includegraphics{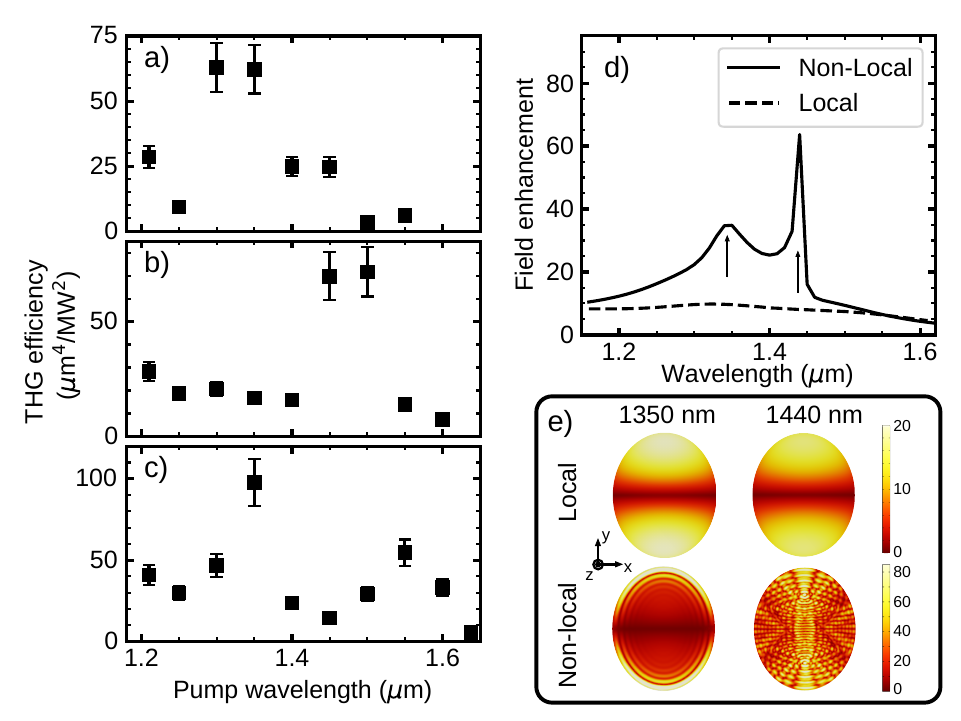}
    \caption{Pump wavelength dependence of the measured instantaneous THG efficiency for an antenna in arrays I (a), II (b) and III (c) (black squares). Error bars represent the effect of compounded uncertainties in the calculated values.
    All curves correspond to the same antennas investigated in Figure\;\ref{fig:dfss}. (d) Volume averaged electric field enhancement inside the ITO for a PGA with the nominal dimensions of array II with a local (dashed line) and nonlocal (full line) dielectric function. Arrows denote the incident wavelengths corresponding to the maps shown in e). (e) Simulated maps of the field enhancement in the middle of the ITO layer of a PGA with the nominal dimensions of array II with a local and nonlocal dielectric function. The profiles are shown for an incident plane wave at 1345\:nm (left column) and 1440\;nm (right column), corresponding to the arrrows in d).
    }
    \label{fig:THGWave}
\end{figure}

To extract the THG power conversion efficiency in a way that is independent of pump irradiance and pulse duration, we compute the ratio of the peak third harmonic irradiance to the cube of the peak pump irradiance. Note that the values reported here strongly underestimate the internal THG efficiency due to the high absorption at the third harmonic within the a-Si dielectric and the finite numerical aperture of the collection objective. Simulations shown in the Supporting Information estimate that 98\% of the THG power is absorbed by the a-Si and only 23\% of the radiated THG is collected by our apparatus due to the highly anisotropic THG radiation profile. We measured the instantaneous THG efficiency of antennas within arrays I, II and III using pump wavelengths spanning 1.21\;$\mu$m to 1.64\;$\mu$m. The corresponding values are shown in Figure\;\ref{fig:THGWave} a) to c).
The PGAs chosen for this investigation are the same as those whose DFS spectra are plotted in Figure \ref{fig:dfss}. For all antennas, clear resonances are observed atop a broad THG response within the measured wavelengths. Despite showing strong THG over their whole field enhancement band, the trends followed by the THG data differ qualitatively from the calculated field enhancement spectra cubed (see Fig.~\ref{fig:fieldenh}). In addition to the unexpected presence of sharp resonances, the highest enhancement (see scale) is observed for the widest antenna, in contrast to the theoretical prediction. Several mechanisms can contribute to these differences. In particular, the a-Si absorption nearly triples over the probed third harmonic spectral window of 550\;nm to 400\;nm\cite{pierce_electronic_1972}, which significantly reduces the radiated THG power at shorter wavelengths. The ITO's nonlinear susceptibility itself has a frequency dependence that is unaccounted for in a simple field enhancement calculation: the longitudinal and transverse components of its nonlinear susceptibility have been previously reported to increase by approximately $60\%$ and $100\%$, respectively, in the range between 1020\;nm to 1320\;nm\cite{rodriguez-sune_retrieving_2021} and the contribution from THG generated from a-Si is ignored in this simple picture. Finally, our calculations ignore nonlocal effects, which due to the nanometric thickness of the ITO layer can be expected to play an important role in the generation and coupling of localized THG to far-field radiation. The first three mechanisms are expected to lead to quantitative differences, but cannot explain the appearance of sharp resonances.

To better understand the near-field optical properties of the PGAs, we incorporate nonlocal effects into the dielectric function of the ITO using the same approach as for DFS. By restraining our discussion to the field enhancement only, we neglect nonlocal contributions to the ITO's non-linear permittivity (and polarization), which would be required for a more detailed description.
In Figure \ref{fig:THGWave} d), we show the volume averaged field enhancement within the ENZ layer of a PGA of array II as a function of incident wavelength, assuming a local (dashed line) and nonlocal (solid line) ITO permittivity. In contrast to its local counterpart, the nonlocal spectrum presents resonant maxima atop a broad background. To explore the origin of these nonlocal peaks, Fig.~\ref{fig:THGWave} e) shows field enhancement maps within a plane taken in the middle of the ITO layer at $\lambda=1350$~nm and $\lambda=1440$~nm, the wavelengths indicated by vertical arrows in Fig.~\ref{fig:THGWave} d). As expected, the local simulations show smooth, dipolar-like profiles with a single node along the minor axis of the PGA's cross-section. On the contrary, the nonlocal maps exhibit tight oscillatory modulations, atop a dipolar-like background, clearly apparent at 1350 nm. In agreement with the volume averaged spectra in Fig.~\ref{fig:THGWave} d), the maximum field enhancement in the nonlocal maps is a factor $\sim3$ larger than in the local ones (note the color scales).

As discussed in the Supporting Information, the origin of the nonlocal spectral peaks and fast field spatial oscillations in Figs.~\ref{fig:THGWave} d) and e), respectively, can be linked to the positive asymptotic slope that the dispersion of the long-range surface plasmons supported by the ITO film acquires when a hydrodynamic description of the permittivity is used. This is shown in Fig. S10, which also includes the flat asymptote characteristic of the local description. Thus, we can associate the peak at 1350 nm with the zero-group velocity point that the dispersion relation presents at low $k$, obtained under both descriptions. This is accompanied by high wavevector modes only in the nonlocal case, which are responsible for the fast oscillations in the corresponding field map. The peak at 1440 nm emerges due to the second, higher $k$, zero-group velocity peak in the nonlocal band (absent in the local picture). The field enhancement in this case reflects the excitation of tightly confined, long-range surface plasmons within the ITO layer by the incident fields, which explains the nanometric size of the features apparent in the color map.
Our full-wave simulations indicate that changing the cross-sectional dimensions of the PGA does not significantly shift these resonances (see Figure S11 of Supporting Information). We find, however, that the frequency of the high $k$ zero group velocity point shifts considerably with minute changes in the thickness of the ITO layer. This is consistent with the strong sensitivity of the THG efficiency to the ITO microstructure and the much larger standard deviation between nominally identical antennas, shown as error bars in Fig.~\ref{fig:THGWave} a-c), at the wavelengths of peak efficiency. Instead of relying on field enhancement as a proxy for THG, a more complete description could be obtained by performing a full-wave calculation of the third harmonic polarization in the presence of nonlocality. This, however, is complex to implement \cite{Scalora2020}.

In light of our electromagnetic simulations, we interpret the resonances in THG efficiency observed in Figure \ref{fig:THGWave} a) to c) to arise from the contribution of nonlocal effects to the field enhancement. This would imply that the shifts of the THG efficiency resonance positions between arrays with different PGA cross-sections is extrinsic and could be due to a slight structural variations in ITO layer thickness our doping density.  For comparison's sake, absolute THG power conversion efficiencies measured for the antennas studied in Figure\;\ref{fig:THGWave} are 8.6$\times 10^{-8}$ at 3.7 GW/cm$^2$, 4.7$\times 10^{-8}$ at 2.6 GW/cm$^2$ and 5.1$\times 10^{-8}$ at 3.2 GW/cm$^2$ for arrays I, II and III excited at 1.35, 1.45 and 1.55 $\mu$m respectively. These figures include the large absorption losses in the a-Si and the limited collection efficiency. From simulations of the THG external coupling efficiency, we expect internal efficiencies to be 200 times larger than the figures reported here. Given that the dominant loss mechanism is simply re-absorption of the THG radiation within the a-Si, moving to longer wavelengths or a transparent dielectric could alleviate these losses.

\section{Conclusion}

We have demonstrated photonic gap antennas in which an ultra thin layer of ITO, an ENZ material, is embedded.
By comparing dark field scattering spectroscopy measurements with simulations of the antennas' scattering cross-section, we showed that they host hybrid polaritonic modes in accordance with simulations and previous work. Through THG, we confirm that these modes tightly confine the fields inside the 10\;nm thick ITO layer, leading to high field enhancement over a broad spectral range of 400\;nm. This proof of concept demonstrates the effectiveness of PGAs as vertically stacked optical antennas. Although the antenna geometries were chosen to give a high field enhancement at the pump wavelength, they were not specifically optimized for THG. The THG efficiency could be significantly improved by moving to dielectric material that is transparent at the THG wavelength and adjusting the antenna dimensions to ensure efficient radiation of the third harmonic polarization.  Certain experimental features of this work cannot be understood within the framework our local full-wave calculations. Namely unexplained spectral shifts and variations in the relative amplitude of the DFS peaks and resonances occurring atop the broadband THG response. We find that including nonlocality in the dielectric response of ITO used in our model improves the agreement for DFS and leads to the appearance of sharp resonances in the field enhancement spectra, both of which improve the agreement with experiment. An important point is that the inclusion of nonlocality leads to larger field enhancement maxima than those calculated in a local model. These maxima can be linked to the presence of points of vanishing group velocity in the long-range surface plasmon dispersion relation of the ITO layer when nonlocal corrections are included. In conclusion, PGAs afford strong localized field enhancement in an device that can be fabricated using photolithography or straightforward electron-beam lithography, forgoing the high lateral resolution requirements of traditional antenna designs. For thin ENZ layers, nonlocal corrections must be included in the model. They contribute to both a larger field enhancement values and the appearance of qualitatively different field enhancement patterns.
While our demonstration is in the near-infrared, PGAs are particularly attractive for applications in the mid-infrared where the ENZ frequency of doped III-V semiconductors is located\cite{jun_doping-tunable_2014}.
Using molecular beam epitaxy, infrared ENZ materials could be grown inside PGAs with deep subwavelength thicknesses well below those reported here and provide more than an order of magnitude larger field enhancement. We also wish to highlight that during revision of this manuscript, another experimental study~\cite{Tirole2024} was published on THG in ENZ PGAs comprising thicker 20~nm ITO layers, which did not report a contribution from nonlocal effects.

\section{Acknowledgements}
This work was supported by the Natural Sciences and Engineering Council of Canada Strategic Grant Program, the Canada Research Chairs Program and the joint call with the European Union's Horizon Europe Research and Innovation Programme under agreement 101070700 (MIRAQLS). F.T. acknowledges financial support from the Canada Excellence Research Chair (CERC-2022-00055). We would also like to acknowledge design and fabrication support from CMC Microsystems and Canada’s National Design Network (CNDN).

\subsection{Author Contributions}
* F.T, D.M.M and A.P have contributed equally to this work.

\section{Supporting Information}
Ellipsometry and structural details, additional dark field scattering and third harmonic generation data, further experimental setup details, simulated third harmonic outcoupling, nonlocal dispersion relations and calculation details.

\section{Methods}
\subsection{Sample fabrication}

A B270 glass substrate (51\;mm x 51\;mm and 2\;mm thick) was pre-cleaned using an oxygen and argon plasma.
The PGAs were fabricated by sputtering in a CMS-18 deposition system by Kurt J. Lesker using 3-inch targets of Si and ITO (In2O3/SnO2 90/10 wt \%). Each layer of material was grown on a glass substrate to the desired thicknesses.
A 372\;nm layer of amorphous silicon was then sputtered at 200\;\degree C, followed by a 10\;nm layer of ITO sputtered at room temperature.
The sample was then removed from the chamber and measured on a J.A. Woollam RC2-XI ellipsometer.
After returning it to the sputtering chamber, it was annealed under vacuum for an hour at a 400\;\degree C setpoint in order to reduce the epsilon-near-zero wavelength to 1188\;nm.
The temperature was then brought down to 200\;\degree C for the sputtering of the final 128\;nm layer of amorphous silicon.
The layer thicknesses after annealing and bulk optical properties were then re-characterized in situ using a M-2000 ellipsometer (see Figure S1 of Supporting Information).

The large sample was cleaved into several smaller samples roughly equal in size, which were cleaned by sonication in acetone and isopropyl alcohol followed by ten minutes in an oxygen plasma (200\;W, 0.8\;mbar).
Immediately before spin coating, each sample was vacuum baked (50\;torr, 120\;\degree C, five minutes) and primed with SurPass 3000.
They were spin-coated with MA-N 2403 (4000 rpm) negative electron beam resist, pre-baked on a hotplate (90\;\degree C, 1 minute), and exposed in a Raith 150 electron beam lithography system (20\;kV, 80\;$\mu$C/cm$^2$).
Development was done in MA-D 525 developer (40 seconds) followed by two rinses in deionized water while stirring (10 minutes total) then by a post-bake on a hotplate (90\;\degree C, 2 minutes).
The samples were then mounted onto an \ch{Al2O3}-coated silicon carrier wafer using Crystalbond and etched in an Oxford Instruments Plasmalab 100 ICP 180 reactive ion etching tool.
The recipe for etching the top amorphous silicon (1200\;W ICP power, 60\;W RF, 49\;V bias, 20\;mtorr, 15\;\degree C, 22\;sccm \ch{SF6}, 38\;sccm \ch{C4F8}, 16\;seconds) is a pseudo-Bosch process similar to those used elsewhere for anisotropic etching of silicon \cite{con_nanofabrication_2014,saffih_fabrication_2014,khorasaninejad_enhanced_2012,ayari-kanoun_silicon_2016,henry_alumina_2009}.
The sample was removed from the chamber and wet etched in an acid solution (1:3:4 \ch{HNO3}/HCl/\ch{H2O}) for 2 minutes to remove the ITO layer.
The sample was reinserted in the chamber and etched a second time using the same peusdo-Bosch recipe for 35 more seconds, and finally cleaned in an oxygen plasma (2000\;W ICP, 0\;W RF, 35\;mtorr, 15\;\degree C, 30 seconds).

\subsection{Optical characterization}

The optical characterization setup used was designed to investigate both dark field scattering (DFS) and third harmonic generation (THG) in single antennas.
This allows the direct correlation of individual DFS and THG spectra despite possible inhomogeneities between pillar specimens.
A detailed diagram of this setup is presented in the Supporting Information.
The sample was held in an inverted microscope customized to allow illumination straight from the top through a 10x microscope objective for THG or at a grazing angle through a 10\;mm biconvex lens for DFS spectroscopy.
A stabilized laser lamp provided a broad and stable infrared illumination for the DFS experiments. The output of an optical parametric amplifier (200\;fs pulse duration, 100\;kHz repetition rate, Light Conversion ORPHEUS-F in long pulse mode) was used to pump the THG.
For both experiments, light was collected from the bottom of the microscope through a 40x 0.75 NA objective and sent to an infared spectral imager for DFS scattering spectroscopy or a visible light imaging system for THG.
The collected light was directed to either systems using an accurate magnetically registered kinematic mount.
The visible imager consists of a single lens, a visible shortpass filter (400\;nm to 550\;nm bandpass, Thorlabs FESH550) and a silicon CCD camera (Princeton Instruments Pixis 400) sensitive from 350\;nm to 1000\;nm.
The infrared imager is composed of a InGaAs CCD camera (First Light C-RED2) sensitive from 800\;nm to 1650\;nm connected at the exit of an imaging spectrometer.

When performing DFS experiments, only the infrared imager was used with the spectrometer's first diffractive order. The spectrum of the broadband light source was obtained by focusing it at the sample plane from within the collection objective's numerical aperture and recording it using the infared imager.
The spectrum of the DFS cross-section was obtained by dividing the scattered spectrum by that of the light source after proper background subtraction.

During THG measurements, the pump profile at the sample was first imaged using the infrared imager with its spectrometer at the zeroth order and a neutral density filter. The pump was then turned off with a mechanical shutter and the sample's region of interest was moved under the pump's illumination. The collection path is then reconfigured for the use of the visible imager and the pump shutter opened to acquire an image of the pump induced THG. Using a ruled sample, the infrared and visible imagers' image planes are calibrated so that a location on one of these planes can be mapped onto the other. This allows for the measurement of the pump power density responsible for THG in a given pillar.

All pump powers were measured with a high resolution thermal power sensor (Thorlabs S401C). The THG power was measured by calibrating the number of counts on the visible imager's camera to the power emitted at the sample's plane. This calibration was performed for every wavelength of interest using the same acquisition parameters as those used in the THG measurements.

To obtain the THG efficiency from the average power, we take the pump pulses to be Gaussian with a 200 fs full width at half maximum at all wavelengths and assume that the generated THG pulses are also Gaussian, but with a pulse duration reduced by a factor $\sqrt{3}$, as expected for THG over a bandwidth much greater than that of the pump pulse.

\subsection{Full wave calculations}
Finite-element method calculations of the dark field scattering and field enhancement were performed using COMSOL Multiphysics. In both calculations, the field of the incident plane wave (from the substrate side) was first calculated in the absence of the PGA using periodic boundary conditions on the four side boundaries and Ports at the bottom and top boundaries to account for incoming and outgoing waves, respectively. The result was then used as a background field within the scattered field formulation along with perfectly matched layer boundary conditions to compute the response in the presence of the PGA. For dark field scattering, the plane wave angle of incidence ($56\degree$ in air) and polarization were chosen to match experiment and the scattered power, obtained from the real part of the Poynting vector, was only integrated within a solid angle corresponding to the 0.75 NA of the collection objective at the top boundary (air-side). For field enhancement, the angle of incidence was chosen to be zero.

Non-local simulations were performed by adapting the above geometry using the simple model of Ref.~\citenum{Luo2013}. In this case, a \textit{d} = 1 nm thick shell was included within the ITO layer to reproduce the effects of nonlocality. The shell's  anisotropic index was chosen to be vanishing along the tangential direction ($\epsilon= 0.001$ in the implementation) and following Ref.~\citenum{Luo2013} in the normal direction.

\bibliography{bibliography}
\end{document}